\documentclass[11pt]{article}

\usepackage[margin=1in]{geometry}
\usepackage{amsmath,amssymb,mathtools,algorithm,algorithmic,url,cite,bm,amsthm}
\usepackage{newpxtext,newpxmath}
\usepackage{stmaryrd}
\usepackage[hidelinks]{hyperref}
\usepackage{graphicx}
\usepackage[labelformat=simple]{subcaption}

\usepackage[margin=10pt,font=small,labelfont=bf,labelsep=endash]{caption}

\newtheorem{proposition}{Proposition}
\newtheorem{theorem}{Theorem}

\newcommand\colororbw{color}

\DeclareCaptionStyle{ruled}%
{labelfont=bf,labelsep=endash,strut=off}

\newcommand{\ZZ}{\mathbb Z}
\newcommand{\FF}{\mathbb F}
\newcommand{\RR}{\mathbb R}
\newcommand{\CC}{\mathbb C}

\newcommand\V\mathbf

\newcommand\trp{^T}   
\newcommand\htrp{^H}  

\let\Eq\equiv

\let\conj\overline

\DeclareMathOperator\esc{esc}
\DeclareMathOperator\float{float}
\DeclareMathOperator\diag{diag}
\DeclareMathOperator*\round{round}

\DeclareMathOperator\RRe{Re}
\DeclareMathOperator\IIm{Im}
\DeclarePairedDelimiter\floor\lfloor\rfloor
\DeclarePairedDelimiter\ceil\lceil\rceil
\DeclarePairedDelimiter\abs\lvert\rvert

\DeclarePairedDelimiterX{\set}[2]{\lbrace}{\rbrace}{#1 \,\delimsize\vert\, #2}

\newcommand\algref[1]{(line~\ref{#1})}

\newcommand\SYRK{\texttt{SYRK}}

\newcommand\HERK{\texttt{HERK}}

\title{The 2M Multiplication Algorithm for Complex Matrices}
\author{Peter Caday\\\small NVIDIA Corporation\\[-0.4ex]\small\texttt{pcaday@nvidia.com}}

\begin{document}

\maketitle

\begin{abstract}
Complex matrix multiplication is typically computed using 4 real matrix multiplications (GEMMs) of the same size. The well-known 3M multiplication algorithm reduces this cost to 3 real GEMMs, together with quadratic time pre- and post-processing steps. In this paper, we reduce 3M to 2M for matrices with integer real and imaginary parts, performing complex GEMM with only 2 real GEMMs of the same size, along with quadratic time pre- and post-processing. For floating-point matrices, 2M multiplication combines naturally with the Ozaki-II scheme, yielding a practical, high-performance algorithm for computing a complex floating-point GEMM in roughly twice the time of a real GEMM of the same size. As corollaries, we derive new algorithms for symmetric rank-$k$ updates (\SYRK/\HERK) that internally use full rectangular GEMMs.
\end{abstract}


\section{Introduction}

The classical approach to multiply complex matrices $A=A_r+iA_i\in\CC^{m\times k}$ and $B=B_r+iB_i\in\CC^{k\times n}$ is to expand the product into real matrix multiplications:
\[
C = AB = (A_rB_r-A_iB_i) + i(A_rB_i+A_iB_r).
\]
Let us call this the \emph{4M} algorithm, following~\cite{BLISComplex}, since it involves 4 real matrix multiplications. 4M multiplication typically forms the backbone of complex floating-point matrix multiplications (GEMMs) in high-performance libraries.

The well-known \emph{3M} algorithm reduces this cost to 3 real matrix multiplications, by first computing $A_r+A_i$ and $B_r+B_i$:
\begin{equation}
C = \big(A_rB_r - A_iB_i\big) + i\big((A_r+A_i)(B_r+B_i) - A_rB_r - A_iB_i\big).
\label{e:3m}
\end{equation}
Note that while~\eqref{e:3m} is mathematically exact, in floating-point arithmetic the additions $A_r+A_i$, $B_r+B_i$ can reduce numerical stability~\cite{Higham3M}. Hence, numerical libraries typically offer 3M GEMMs as an opt-in feature.

For scalar complex products, it can be proven that 3 real multiplications is indeed minimal~\cite{FZ,Winograd2x2}. In this paper, we will show that we can do better for matrix products.

\section{2M Complex Multiplication}

To start, we consider input matrices $A=A_r+iA_i$, $B=B_r+iB_i$ with integral real and imaginary parts, and their product $C=AB$. Suppose $m$ is an integer chosen large enough that $\abs{\RRe(c_{ij})}, \abs{\IIm(c_{ij})} \leq \frac m2$ for each entry $c_{ij}$ of $C$. Then, if we know $C\bmod m$, we can evidently recover $C$.

Now further suppose that $m$ is odd and there is an integer $s$ such that $s^2\Eq-1\pmod m$, in effect a ``real surrogate'' for $i$. It is well-known from number theory that such an $s$ exists iff $m$ is a product of primes congruent to 1 mod 4~\cite[Thm.\ 80]{HardyWright}. We call such an $m$ a \emph{2M modulus}.

To multiply $AB$, first perform the 3M-like preprocessing steps
\begin{subequations}
\begin{equation}
\begin{aligned}
	A_+ &\Eq A_r + sA_i \pmod m, \hspace{7em} &  B_+ &\Eq B_r + sB_i\pmod m,\\
	A_- &\Eq A_r - sA_i \pmod m, &  B_- &\Eq B_r - sB_i\pmod m.
\end{aligned}
\label{e:2m-pre}
\end{equation}
Next, perform the 2 real multiplications
\begin{align}
	C_+ &\Eq A_+ B_+ \pmod m,
	&
	C_- &\Eq A_- B_- \pmod m.
	\label{e:2m-mul}
\end{align}
Then we will see shortly that
\begin{align}
	C = AB \Eq 2^{-1} \big(C_- + C_+ \big)+2^{-1}s\big(C_- - C_+\big)i\pmod m,
	\label{e:2m-post}
\end{align}
\end{subequations}
where $2^{-1}=(m+1)/2$ is the inverse of 2 modulo $m$.

Together, equations~(\ref{e:2m-pre}--\ref{e:2m-post}) form the \emph{2M multiplication algorithm}. As with 3M, the new algorithm requires $\Theta(mn+mk+nk)$ scalar operations in addition to the matrix multiplications. We start by proving correctness.

\begin{proposition}
\label{p:2m-integer-correctness}
Equation \eqref{e:2m-post} holds for every $A$, $B$.
\end{proposition}
\begin{proof}[Proof I]
Expanding~\eqref{e:2m-mul} yields
\begin{align*}
C_\pm &\Eq A_rB_r \pm s(A_rB_i + A_iB_r) + s^2 A_iB_i \\
	&\Eq \big(A_rB_r - A_iB_i\big) \pm s(A_r B_i+A_iB_r) \pmod m.
\shortintertext{Hence}
C_+ + C_- &\Eq 2\big(A_rB_r - A_iB_i\big) \pmod m,\\
C_+ - C_- &\Eq 2s\big(A_rB_i+A_iB_r\big) \pmod m.
\end{align*}
Both 2 and $s$ are invertible modulo $m$, and $s^{-1}\Eq -s\bmod m$, from which~\eqref{e:2m-post} follows immediately.
\end{proof}

Proof I, while valid, is not very illuminating. In the next section, we will invoke some basic algebraic number theory in order to better motivate 2M multiplication and provide a second, deeper proof of Proposition~\ref{p:2m-integer-correctness}.

\subsection{Algebraic Underpinnings}
\label{s:2m-algebra}

Algebraically, $A$, $B$, $C$ are matrices over the ring of Gaussian integers $\ZZ[i]=\set{x+yi}{x,y\in\ZZ}$. Gaussian integers have a rich algebraic structure, and in particular, they form a Euclidean domain, meaning that there is a well-defined notion of division with remainder. Modular arithmetic and the Chinese remainder theorem (CRT) are likewise valid in $\ZZ[i]$~\cite{HardyWright}.

The 2M algorithm then follows from two observations:
\begin{itemize}
\item First, a 2M modulus $m$ factors into a product of Gaussian integers $w\overline w$. Hence we can compute $AB\bmod m$ by first evaluating $AB\bmod w$ and $AB\bmod{\conj w}$, then applying CRT. 
\item Second, there is an isomorphism $\ZZ[i]/w\ZZ[i]\cong \ZZ/m\ZZ$. Under this isomorphism, the complex residues $A,\,B \bmod w,\,\conj w$ become real-valued, and the complex products $AB\bmod w$ and $AB\bmod{\conj w}$ can be re-expressed as real multiplications mod $m$.
\end{itemize}

\begin{proof}[Proof II of Proposition~\eqref{p:2m-integer-correctness}]
Since $m\Eq 1\bmod 4$ there exist positive coprime $u, v\in\ZZ$ such that $m = u^2+v^2$ (cf.~\cite[Thm.\ 366]{HardyWright}). Letting $w=u+vi\in\ZZ[i]$, it follows that $m=w\conj w$. Note that $u^2v^{-2}\Eq -1\bmod m$, so $s\Eq \pm uv^{-1}\bmod m$. By negating $v$ if necessary, we can force $s\Eq -uv^{-1}\bmod m$.

Next, we note that the mapping $\varphi_+\colon i\mapsto s$ induces an isomorphism of quotient rings $\ZZ[i]/w\ZZ[i]\cong \ZZ/m\ZZ$. Indeed, as a homomorphism $\varphi_+$ is well-defined since $s^2\Eq -1\bmod m$, and it has a trivial inverse $\ZZ/m\ZZ\hookrightarrow \ZZ[i]/m\ZZ[i]\to\ZZ[i]/w\ZZ[i]$, well-defined as $w\mid m$. Similarly, the mapping $\varphi_-\colon i\mapsto -s$ is an isomorphism from $\ZZ[i]/\conj w\ZZ[i]$ to $\ZZ/m\ZZ$.

We can now interpret the 2M algorithm in the light of these isomorphisms. Clearly, $A_\pm = \varphi_\pm(A)$, $B_\pm = \varphi_\pm(B)$, so the preprocessing steps~\eqref{e:2m-pre} are equivalent to reducing $A$, $B$ modulo $w$, $\conj w$ and applying $\varphi_\pm$. Next, the multiplication steps~\eqref{e:2m-mul} compute
\begin{align*}
C_+&=\varphi_+(A\bmod w)\cdot\varphi_+(B\bmod w)=\varphi_+(C\bmod w),\\
C_-&=\varphi_-(A\bmod {\conj w})\cdot\varphi_-(B\bmod {\conj w})=\varphi_-(C\bmod{\conj w}).
\end{align*}
More simply, since the $\varphi_\pm^{-1}$ are inclusions, $C_+\Eq C\bmod w$ and $C_-\Eq C\bmod{\conj w}$.

Finally,~\eqref{e:2m-post} proves to be a specialization of the Chinese Remainder Theorem in $\ZZ[i]$, reconstructing $C\bmod m$ from $C\bmod w$ and $C\bmod{\conj w}$. Namely, the CRT states
\begin{align*}
C&\Eq \chi_+\conj wC_+ + \chi_-wC_-\pmod m,
\end{align*}
where $\chi_+ \Eq \conj w^{-1}\bmod w$, $\chi_- \Eq w^{-1}\bmod \conj w$. We have
\[
\chi_+ \Eq (u-vi)^{-1} \Eq (2u)^{-1}\pmod w
\quad
\Rightarrow
\quad
\chi_+\conj w \Eq 2^{-1}u^{-1}(u - vi) \Eq 2^{-1}(1 - is)\pmod m.
\]
Similarly $\chi_- w = 2^{-1}(1 + is)\bmod m$, establishing~\eqref{e:2m-post}.
\end{proof}

\subsection{Non-2M Moduli}

We conclude by noting a negative result: no 2M-like algorithm is possible for non-2M odd prime moduli. However, the 3M method~\eqref{e:3m} is still valid, being applicable in any ring, and is therefore optimal.

\begin{proposition}
If $m\Eq 3\bmod 4$ is prime, there are no constants $p_{ij}$, $q_{ij}$, $r_{ij}$ such that
\begin{align}
\label{e:would-be-2m}
AB &\Eq \big(C_1r_{11}+C_2r_{21}\big) + \big(C_1r_{12}+C_2r_{22}\big)i \pmod m
\end{align}
holds for all $A$, $B$, where
\begin{align}
\label{e:would-be-2m-2}
C_j &= A_jB_j,
&
A_j &= p_{i1}A_r + p_{i2}A_i,
&
B_j &= q_{i1}B_r + q_{i2}B_i.
\end{align}
\end{proposition}
\begin{proof}
This is an immediate corollary of the general multiplicative complexity results of Fiduccia and Zalcstein~\cite{FZ}.
Because $-1$ is not a quadratic residue for any $m\Eq 3\pmod 4$, the polynomial $s^2+1$ is irreducible modulo $m$, and $\ZZ[i]/m\ZZ[i]\cong(\ZZ/m\ZZ)[s]/(s^2+1)\cong \FF_{m^2}$ is a degree-2 extension of $\ZZ/m\ZZ$. Theorem 8 in~\cite{FZ} then states that multiplication in $\ZZ[i]/m\ZZ[i]$ can be performed by 3 multiplications in $\ZZ/m\ZZ$ by non-constants, and no fewer. In particular, it is impossible to choose $p_{ij}$, $q_{ij}$, $r_{ij}$ to make~\eqref{e:would-be-2m}--\eqref{e:would-be-2m-2} true, even in the special case where the input matrices $A$, $B$ are scalars.
\end{proof}

\section{2M Multiplication In Floating Point}

\subsection{The Ozaki-II Scheme}

To apply 2M to non-integer GEMMs, we employ the Ozaki-II scheme~\cite{Ozaki2, Ozaki2_2}, a family of algorithms for computing floating-point GEMMs to arbitrary accuracy. Ozaki-II works by scaling and quantizing its inputs to integers, then performing fast integer matrix multiplication with modular arithmetic. By leveraging the dedicated low-precision matrix multiplication units available on modern GPUs, Ozaki-II can often significantly outperform standard floating-point GEMM while maintaining comparable accuracy. Moreover, the level of accuracy may be tuned: either reduced, for additional performance; or increased, to provide additional accuracy beyond that available with a standard floating-point implementation.

We briefly restate the method here. Given real matrices $\mathbf A=(\mathbf a_{ih})\in\RR^{m\times k}$, $\mathbf B=(\mathbf b_{hj})\in\RR^{k\times n}$, we choose per-row and per-column scaling factors $\sigma\in\RR^m$, $\tau\in\RR^n$ for $\mathbf A$ and $\mathbf B$, respectively, and quantize to integer matrices
\[
A = \big(\round_\ZZ(\mathbf a_{ih}^{\vphantom1}\sigma_i^{-1})\big)\in\ZZ^{m\times k},\qquad
B = \big(\round_\ZZ(\mathbf b_{hj}^{\vphantom1}\tau_j^{-1})\big)\in\ZZ^{k\times n}.
\]
Here, $\round_S(\cdot)$ represents rounding to a nearest%
	\footnote{In the original Ozaki-II implementation, round-to-zero is used instead of round-to-nearest~\cite{Ozaki2_2}. Rounding to nearest provides an additional bit of accuracy, but any rounding is valid.} %
element of a set $S$. Throughout, we employ boldface letters to denote real- and complex-valued variables and italics for their quantized integer counterparts.

Next, we choose a collection of coprime moduli $m_1,\dotsc,m_N\in\ZZ_+$ whose product $M$ is large enough that each element of the integer product $C=AB$ has absolute value at most $\frac M2 - 1$. Then $C=(c_{ij})$ can be uniquely recovered from its residue modulo $M$, and we do so by computing the product residues
\begin{equation}
C_\ell \Eq \left(A\bmod m_\ell\right)\left(B\bmod m_\ell\right)\pmod{m_\ell},\qquad \ell=1,\dotsc,N,
\label{e:o2-products}
\end{equation}
and applying the Chinese remainder theorem (CRT) elementwise:
\begin{equation}
C \Eq \sum_{\ell=1}^N C_\ell\cdot\left(\frac{M}{m_\ell} \bmod m_\ell\right)^{-1} \frac{M}{m_\ell} \quad\pmod M.
\label{e:crt}
\end{equation}
Finally, we re-apply the scaling factors to reconstruct an approximation $\tilde{\mathbf C}=(c_{ij}\sigma_i\tau_j)$ to the exact product $\mathbf C=(\mathbf c_{ij})$.

Ozaki-II derives its speed from computing the products~\eqref{e:o2-products} using dedicated low-precision matrix multiplication hardware, which guides the choice of the moduli $m_\ell$ (e.g., $m_\ell\leq 2^8$ for 8-bit integer GEMMs).

The accuracy of the Ozaki-II method, much like the accuracy of floating-point GEMMs, is an involved topic and depends strongly on the input data, as well as the scaling factors $\sigma$, $\tau$~\cite{O1Error,O2Error,ESC,Abdelfattah,FMM}. However, for \emph{fixed} inputs $\mathbf A$, $\mathbf B$, elementwise error $\abs{\textbf c_{ij} - c_{ij}\sigma_i\tau_j}$ scales as $O(M^{-1/2})$ for properly chosen $\sigma$, $\tau$, so it is often useful to view $M$ as an accuracy parameter. We will study accuracy and the choice of $\sigma$, $\tau$ in more detail in section~\ref{s:accuracy}.

\subsection{Complex Ozaki-II With 2M}
\label{s:2m-ozaki2}

Ozaki-II extends readily to complex-valued inputs $\mathbf A=\mathbf A_r+i\mathbf A_i\in\CC^{m\times k}$, $\mathbf B = \mathbf B_r+i\mathbf B_i\in\CC^{k\times n}$. In Uchino et al.'s algorithm~\cite{Uchino}, we first choose (real) scaling factors $\sigma$, $\tau$ and quantize $\mathbf A_r$, $\mathbf A_i$, $\mathbf B_r$, $\mathbf B_i$ to integer matrices as above. The resulting integer matrices are reduced modulo each $m_\ell$, multiplied, and then the real and imaginary parts of $C$ are reconstructed separately using CRT. Scaling factors may be chosen using straightforward extensions of any of the existing real algorithms~\cite{Ozaki2_2,ESC} or based on a desired integer bit width~\ref{e:standard-sf}.

For better performance, Uchino et al.\ apply the 3M method to each small integer multiplication, after modular reduction%
\footnote{%
It is also possible to apply 3M \emph{prior} to modular reduction, but this has the disadvantage of requiring high-precision integer arithmetic in 3M's pre- and post-processing steps.}%
; since 3M is an algebraic identity, it is valid modulo any $m_\ell$. By substituting 2M multiplication for 3M, we immediately derive a 2M Ozaki-II algorithm for complex floating-point GEMMs (Algorithm~\ref{a:2m-ozaki2}). 

\begin{algorithm}[t]
\caption{2M Ozaki-II}
\begin{algorithmic}[1]
\STATE Fix a set of coprime 2M moduli $m_1,\dotsc,m_N$; let $M=\prod_{\ell=1}^N m_\ell$.
\STATE Choose scaling factors $\sigma\in\RR^{m}$, $\tau\in\RR^{n}$.		\label{a:2m:sf}
\STATE Quantize input matrices: $A=\big(\round_{\ZZ[i]}(\sigma_i^{-1}\mathbf a_{ih}^{\vphantom1})\big)$, $B=\big(\round_{\ZZ[i]}(\tau_j^{-1}\mathbf b_{hj}^{\vphantom1})\big)$.			\label{a:2m:quant}
\FOR{$\ell=1,\dotsc,N$}
	\STATE Reduce $A$, $B$ modulo $m_\ell$.
	\STATE Compute $C_\ell=C\bmod m_\ell$ with~(\ref{e:2m-pre}--\ref{e:2m-post}).	\label{a:2m:mul}
\ENDFOR
\STATE Reconstruct $C\bmod M$ from $(C\bmod m_1,\dotsc,C\bmod m_N)$ via the Chinese Remainder Theorem.	\label{a:2m:crt}
\STATE Lift $C$ to $\ZZ[i]^{m\times n}$ by choosing representatives with real and imaginary parts in $[-M/2, M/2)$.
\STATE Reconstruct $\tilde{\mathbf C} = (c_{ij}\sigma_i\tau_j)\in \CC^{m\times n}$.
\end{algorithmic}
\label{a:2m-ozaki2}
\end{algorithm}

In Algorithm~\ref{a:2m-ozaki2} we note several points of customization:
\begin{itemize}
\item \emph{Choice of scaling factors~\algref{a:2m:sf}.} $\sigma$, $\tau$ may be chosen using one of Uchino et al.'s scaling algorithms (``fast'' or ``accurate'')~\cite{Uchino}, or based on a fixed integer bit width (equation~\eqref{e:standard-sf}). See Section~\ref{s:accuracy} for more details.
\item \emph{Rounding method~\algref{a:2m:quant}.} The choice of rounding method is not essential to the algorithm. We suggest round-to-nearest for best accuracy, though round-to-zero may also be used, as in~\cite{Ozaki2_2,Uchino}.
\item \emph{Low-precision matrix multiplication~\algref{a:2m:mul}.} Individual matrix products modulo $m_\ell$ may be performed with any available accelerated matrix hardware. We will consider int8, fp8, and fp4 specifically in Section~\ref{s:examples}.
\item \emph{CRT reconstruction algorithm~\algref{a:2m:crt}.} Numerous algorithms are available for CRT reconstruction, including direct reconstruction via~\eqref{e:crt} and mixed-radix conversion~\cite{MRC}. Uchino et al.\ propose direct reconstruction in double-double arithmetic, though this loses precision for small entries of $C$, when $\log_2 \abs{c_{ij}/M}\lesssim -53+\ceil{\log_2 N}$.
\end{itemize}

We may also readily hybridize the 2M and 3M Ozaki-II methods. Algorithm~\ref{a:hybrid-2m-3m} presents a hybrid approach that supports any set of coprime moduli, using 2M multiplication for moduli $m_\ell$ that satisfy the 2M conditions (odd, with a square root of $-1$), and 3M multiplication for the remaining moduli. For a practical implementation, this hybrid approach can attain higher precision (larger $M$) than Algorithm~\ref{a:2m-ozaki2}, since more moduli are available. In section~\ref{s:examples}, we will discuss how to pick modulus sets in Algorithm~\ref{a:hybrid-2m-3m} for best performance; generally speaking, we prefer 2M moduli to non-2M moduli, but there are several subtleties.

\begin{algorithm}[ht]
\caption{Hybrid 2M/3M Ozaki-II}
\begin{algorithmic}[1]
\STATE Fix a set of arbitrary coprime moduli $m_1,\dotsc,m_N$; let $M=\prod_{\ell=1}^N m_\ell$.
\STATE Choose scaling factors $\sigma\in\RR^{m}$, $\tau\in\RR^{n}$.
\STATE Quantize input matrices: $A=\big(\round_{\ZZ[i]}(\sigma_i^{-1}\mathbf a_{ih}^{\vphantom1})\big)$, $B=\big(\round_{\ZZ[i]}(\tau_j^{-1}\mathbf b_{hj}^{\vphantom1})\big)$.
\FOR{$\ell=1,\dotsc,N$}
	\STATE Reduce $A$, $B$ modulo $m_\ell$.
	\IF{$m_\ell$ is a 2M modulus}
		\STATE Compute $C_\ell = C\bmod m_\ell$ with~(\ref{e:2m-pre}--\ref{e:2m-post}).
	\ELSE
		\STATE Compute $C_\ell$ with~(\ref{e:3m}).
	\ENDIF
\ENDFOR
\STATE Reconstruct $C\bmod M$ from $(C\bmod m_1,\dotsc,C\bmod m_N)$ via CRT.
\STATE Lift $C$ to $\ZZ[i]^{m\times n}$ by choosing representatives with real and imaginary parts in $[-M/2, M/2)$.
\STATE Reconstruct $\tilde{\mathbf C} = (c_{ij}\sigma_i\tau_j)\in \CC^{m\times n}$.
\end{algorithmic}
\label{a:hybrid-2m-3m}
\end{algorithm}

\subsection{Algebraic Variants of 2M Ozaki-II}
\label{s:2m-ozaki2-algebra}

As before, we can obtain additional insight into 2M Ozaki-II by further examining the algebra.
After quantization, Algorithm~\ref{a:2m-ozaki2} is presented as a two-level method. At the top level we reduce integer inputs $A$, $B$ modulo each real $m_\ell$, multiply modulo $m_\ell$ with 2M, and reconstruct the product $C$ with CRT. 
At the lower level, as shown in Section~\ref{s:2m-algebra}, the 2M multiplication steps further factor each $m_\ell$ as a product $w_\ell\conj{w_\ell}$, reduce modulo $w_\ell$ and $\conj{w_\ell}$, multiply, and apply CRT.

Hence, we may reinterpret Algorithm~\ref{a:2m-ozaki2} as \emph{the usual Ozaki-II scheme operating over the Gaussian integers, with conjugate pairs of complex moduli}. Similarly, Uchino et al.'s 3M algorithm reduces to Ozaki-II over the Gaussian integers, but with \emph{real} moduli, while hybrid 2M/3M Ozaki-II (Algorithm~\ref{a:hybrid-2m-3m}) uses both real and complex moduli.

This algebraic perspective is useful because it allows us to easily derive additional Ozaki-II variants.
In particular, 2M multiplication and Algorithms~\ref{a:2m-ozaki2} and~\ref{a:hybrid-2m-3m} have direct analogues where we replace $\ZZ[i]$ by another imaginary quadratic number ring. A natural choice is the Eisenstein integer ring $\ZZ[e^{2\pi i/3}] = \set{a+be^{2\pi i/3}}{a,b\in\ZZ}$, which forms a regular hexagonal grid in the complex plane. By varying the number ring, we vary not only the discretization grid, but also the set of valid 2M moduli. Asymptotically, half of all primes are 2M moduli w.r.t.\ any given imaginary quadratic number ring, by Dirichlet's theorem~\cite{NZM}. In practice, however, since moduli are generally bounded by the input range of hardware matrix multiplication units, specific rings may perform slightly better than others, as we will see in Section~\ref{s:zgemm-int8}.

\section{Accuracy of Ozaki-II Complex GEMMs}
\label{s:accuracy}

In this section, we investigate in more detail the accuracy characteristics of Algorithms~\ref{a:2m-ozaki2} and~\ref{a:hybrid-2m-3m}. Before starting, we make two notes:

\begin{enumerate}
\item[(a)] Since the modular multiplications, whether 2M or 3M, are always performed exactly, Algorithms~\ref{a:2m-ozaki2} and~\ref{a:hybrid-2m-3m} produce identical results to each other and to Uchino et al.'s complex 3M Ozaki-II, when given the same scaling factors $\sigma$, $\tau$ and using the same rounding methods and CRT reconstruction algorithms. Hence, all accuracy results in this section are independent of the choice of complex integer multiplication algorithm.

\item[(b)] We assume that CRT reconstruction is performed exactly, then correctly rounded to floating point. Under this assumption, the sole source of error in Algorithms~\ref{a:2m-ozaki2} and~\ref{a:hybrid-2m-3m} is quantization error in the scaled integer approximations $\mathbf A\approx \diag(\sigma)A$ and $\mathbf B\approx B\diag(\tau)$.
\end{enumerate}

\subsection{Absolute Error}

\subsubsection*{Prior Results for Real-Valued GEMM}
Accuracy of integer approximation algorithms for real-valued GEMMs has been studied by several authors. 
Uchino et al.\ provide absolute error bounds for Ozaki-II~\cite{O2Error}, using an approximate CRT reconstruction, as well for the related Ozaki-I integer approximation algorithm~\cite{O1Error}. Abdelfattah et al.~\cite{Abdelfattah} likewise derive absolute error bounds for Ozaki-I.

The usual starting point is to consider scaling factors $\sigma$, $\tau$ that preserve a given number of bits $q$ in each row of $\mathbf A$ and column of $\mathbf B$%
\footnote{
	For some inputs there are advantages in asymmetric quantization, with $\mathbf A$ and $\mathbf B$ quantized to different bit counts; see~\cite{Abdelfattah,FMM}.
}%
. We first record the maximum elements in each row and column, rounding up to powers of 2:
\begin{align*}
\mathbf a_{i,\text{max}} &= \max_{1\leq h\leq k}\abs{\mathbf a_{ih}},
&
\mathbf b_{\text{max},j} &= \max_{1\leq h\leq k}\abs{\mathbf b_{hj}},
\\
\overline{\mathbf a}_{i,\text{max}} &= 2^{\ceil{\log_2 \mathbf a_{i,\text{max}}}},
&
\overline{\mathbf b}_{\text{max},j} &= 2^{\ceil{\log_2 \mathbf b_{\text{max},j}}},
\end{align*}
and set
\begin{align}
\label{e:standard-sf}
\sigma_i &= 2^{-q} \overline{\mathbf a}_{i,\text{max}},
&
\tau_j &= 2^{-q} \overline{\mathbf b}_{\text{max},j},
\end{align}
resulting in quantized matrices $A$, $B$ with entries in the range $[-2^{q},\, 2^{q}]$. The use of power-of-2 scaling factors both simplifies the implementation and avoids loss of accuracy in the final scaling $c_{ij}\sigma_i\tau_j$. With round-to-nearest, the elementwise quantization errors are bounded by
\begin{align}
\label{e:fxp-input-error}
\abs{\mathbf a_{ih}-\sigma_ia_{ih}} &\leq \frac{\sigma_i}2,
&
\abs{\mathbf b_{hj}-\tau_jb_{hj}} &\leq \frac{\tau_j}2,
\end{align}
from which we can derive~(see \cite[eq.\ (3.1)]{FMM}) the forward error bound%
\footnote{
	There is a factor of 2 difference between~\eqref{e:fxp-fwd-error} and~\cite[eq.\ (3.1)]{FMM} since we use round-to-nearest here, rather than truncation.
}
\begin{align}
\label{e:fxp-fwd-error}
\abs{\mathbf c_{ij}-\sigma_i\tau_jc_{ij}} &\leq 2^{-q}k(1+2^{-q-2}) \overline{\mathbf a}_{i,\text{max}} \overline{\mathbf b}_{\text{max},j}
\end{align}
Since the quantized data $A$, $B$ lies in $[-2^{q},\,2^q]$, the integer product $C$ is bounded elementwise by $2^{2q}k$, providing a lower bound $M_{\text{min}}=2^{2q+1}k+1$ for the product of moduli $M$. Together with~\eqref{e:fxp-fwd-error} this implies that for \emph{fixed} inputs $\mathbf A$, $\mathbf B$, there is an $O(M^{-1/2})$ error dependence:
\begin{equation}
\label{e:fxp-M-error}
\abs{\mathbf c_{ij}-\sigma_i\tau_jc_{ij}} = O(M^{-1/2}).
\end{equation}

\subsubsection*{Complex GEMM}
Handling complex-valued $\mathbf A$, $\mathbf B$ is a straightforward generalization of the real case. Defining $\sigma$, $\tau$ as in~\eqref{e:standard-sf}, the quantization error~\eqref{e:fxp-input-error} must be augmented by a factor of $\sqrt 2$ to account for both real and imaginary parts:
\begin{align}
\label{e:fxp-input-error-cx}
\abs{\mathbf a_{ih}-\sigma_ia_{ih}} &\leq \frac{\sqrt 2}2\sigma_i,
&
\abs{\mathbf b_{hj}-\tau_jb_{hj}} &\leq \frac{\sqrt 2}2\tau_j.
\end{align}
The rest of the derivation in~\cite{FMM} continues as before, yielding the forward error bound
\begin{align}
\label{e:fxp-fwd-error-cx}
\abs{\mathbf c_{ij}-\sigma_i\tau_jc_{ij}} &\leq 2^{-q+1/2}k(1+2^{-q-3/2}) \overline{\mathbf a}_{i,\text{max}} \overline{\mathbf b}_{\text{max},j}.
\end{align}
As in the real case, after $q$ is chosen we have an immediate lower bound $M_{\text{min}}=2^{2q+2}k+1$, with an additional factor of 2 from complex multiplication, and once again~\eqref{e:fxp-M-error} holds.

\subsection{Matching Floating-Point Accuracy}

\subsubsection*{Prior Results for Real-Valued GEMM}
Given the absolute error bound~\eqref{e:fxp-fwd-error}, the natural question is how to choose $q$ to achieve GEMM accuracy comparable to standard floating-point arithmetic. Let $Q$ be the floating-point mantissa width, including the implicit leading 1. Evidently, $q$ should be at least $Q$, but in general additional bits are required, driven by exponent variations within a row of $\mathbf A$ or column of $\mathbf B$~\cite{Abdelfattah}. Schwarz et al.~\cite{ESC} capture this effect more precisely with the \emph{exponent span capacity} (ESC), defined as:
\begin{align*}
\esc(\mathbf A,\mathbf B) &= \max_{i,j}\left[\floor*{\log_2 \max_h \abs{\mathbf a_{ih}}} + \floor*{\log_2 \max_h \abs{\mathbf b_{hj}}} - \floor*{\log_2\max_h \abs{\mathbf a_{ih}\mathbf b_{hj}}}\right].
\end{align*}
Schwarz et al.\ suggested setting $q=Q+\esc(\mathbf A,\mathbf B)+1$. Fasi et al.~\cite[Section 3.1]{FMM} carry the analysis further, proving that by choosing $q=q^*=Q+\esc(\mathbf A,\mathbf B)+3$ the \emph{worst-case} error bound~\eqref{e:fxp-fwd-error} is no larger than Jeannerod and Rump's worst-case floating-point error bound for dot products~\cite{JR}. We recall the latter bound here as we will revisit it in the complex domain:

\begin{theorem}[Jeannerod and Rump~\cite{JR}, Theorem 4.2]
\label{t:jr}
Let $x,\,y\in\FF^k$ be length-$k$ vectors of finite values in a floating point system $\FF$ of mantissa width $Q$ (including leading 1). Let $\float(x\cdot y)\in\FF$ denote the result of evaluating the inner product $x\cdot y$ using a sequence of correctly rounded multiplications and additions, in any order. Then
\begin{equation}
\label{e:jr-dp-bound}
\abs{\float(x\cdot y)-x\cdot y} \leq k2^{-Q}(\abs{x}\cdot\abs{y}),
\end{equation}
assuming no underflow or overflow conditions are encountered, where the absolute values on $x$, $y$ are taken elementwise.
\end{theorem}

It is important to note that using $q=Q+\esc(A,B)+3$ mantissa bits for Ozaki-II does not ensure the result will always have smaller error than a standard floating-point GEMM; for a given choice of input data, Ozaki-II may be more or less accurate than a standard floating-point GEMM. However, Fasi et al.'s analysis does provide a formal justification for using ESC to determine $q$.

\subsubsection*{Complex GEMM}

ESC results for real GEMM readily generalize to the complex domain. The main ingredient is a complex analogue of Jeannerod and Rump's error bound, for standard 4M multiplication.

\begin{proposition}
\label{p:jr-dp-bound-cx}
Let $x=x_r+ix_i$, $y=y_r+iy_i$, where $x_r,\,x_i,\,y_r,\,y_i\in\FF^k$ are length-$k$ vectors with finite values. Let $\hat z=\float(x_r\cdot y_r-x_i\cdot y_i) + i\cdot \float(x_r\cdot y_i+x_i\cdot y_r)$, where $\float(x_1\cdot y_1\pm x_2\cdot y_2)$ denotes the result of evaluating a sum or difference of inner products using a sequence of correctly rounded multiplications and additions, in any order. Then
\begin{align}
\label{e:jr-dp-bound-cx}
\abs{\hat z-x\cdot y} &\leq ck2^{-Q}(\abs{x}\cdot\abs{y}),
&
c &= 2\sqrt 2,
\end{align}
in the absence of underflow and overflow.
\end{proposition}
The constant $c=2\sqrt 2$ is likely not optimal. For the special case $n=1$ it is known that the optimal constants are $c=\sqrt 5$ without fused multiply-add (FMA)~\cite{ComplexFPNoFMA}, and $c=2$ with FMA~\cite{ComplexFPWithFMA}. A proof that $c=2\sqrt 2$ suffices is at the end of the section.

Once we have Proposition~\ref{p:jr-dp-bound-cx}, we may reuse Fasi et al.'s derivation to prove:

\begin{proposition}
\label{p:esc-cx}
If $q\geq q^*= Q+\esc(\mathbf A, \mathbf B) + 2$, then the worst-case error bound~\eqref{e:fxp-fwd-error-cx} for Algorithms~\ref{a:2m-ozaki2} and~\ref{a:hybrid-2m-3m} is no greater than the 4M floating-point worst-case error bound~\eqref{e:jr-dp-bound-cx}.
\end{proposition}

\begin{proof}[Proof of Proposition~\ref{p:esc-cx}]
Let $\hat{\mathbf c}_{ij}$ be the 4M floating-point dot product of the $i^{\text{th}}$ row of $A$ and $j^{\text{th}}$ column of $B$. By Proposition~\ref{p:jr-dp-bound-cx},
\begin{align}
\label{e:fp-dp-bound}
	\abs{\mathbf c_{ij}-\hat{\mathbf c}_{ij}} \leq ck\cdot 2^{-Q} \sum_{h=1}^k \abs{\textbf a_{ih}\textbf b_{hj}}.
\end{align}
Substituting $q=Q+\esc(\mathbf A, \mathbf B) + 2$ into the right-hand side of~\eqref{e:fxp-fwd-error-cx},
\begin{align*}
\abs{\mathbf c_{ij}-\sigma_i\tau_jc_{ij}} 
	&\leq \frac{\sqrt 2}{4}k\cdot2^{-Q}\cdot 2^{-\esc(\mathbf A,\mathbf B)} (1+2^{-q-3/2})\mathbf a_{i,\text{max}} \mathbf b_{\text{max},j}\\
	& \leq \frac{\sqrt 2}{4}k\cdot 2^{-Q} \cdot 2^{-\floor{\log_2 \mathbf a_{i,\text{max}}}}
		2^{-\floor{\log_2 \mathbf b_{\text{max},j}}}
		2^{\floor{\log_2 \max_h \abs{\textbf a_{ih}\textbf b_{hj}}}}
	  (1+2^{-q-3/2})\overline{\mathbf a}_{i,\text{max}} \overline{\mathbf b}_{\text{max},j} \\
	 & \leq \sqrt 2 k\cdot2^{-Q}\cdot \max_h \abs{\textbf a_{ih}\textbf b_{hj}}\cdot (1+2^{-q-3/2}) \\
	 & \leq ck\cdot2^{-Q}\sum_h \abs{\textbf a_{ih}\textbf b_{hj}}.
\end{align*}
This is the desired result. Note that the inequality remains true even if we take the stricter bound $c=2$ in~\eqref{e:jr-dp-bound-cx}.
\end{proof}

\begin{proof}[Proof of Proposition~\ref{p:jr-dp-bound-cx}]
Rewrite $\hat z$ as $\float(X\cdot Y_r)+i\float(X\cdot Y_i)$, where
\begin{align*}
X&=\begin{bmatrix} x_r\\x_i \end{bmatrix},
&
Y_r&=\begin{bmatrix} y_r\\-y_i \end{bmatrix},
&
Y_i&=\begin{bmatrix} y_i\\y_r \end{bmatrix}.
\end{align*}
Applying Theorem~\ref{t:jr} to each real dot product gives
\begin{align*}
\abs{\float(X\cdot Y_r) - \RRe (x\cdot y)} &\leq 2k\cdot 2^{-Q}(\abs{X}\cdot\abs{Y_r}) = 2k\cdot 2^{-Q}\sum_{h=1}^k \left[\abs{x_{r,h}}\abs{y_{r,h}} + \abs{x_{i,h}}\abs{y_{i,h}}\right] \leq 2k\cdot 2^{-Q}(\abs x\cdot \abs y),
\\
\abs{\float(X\cdot Y_i) - \IIm (x\cdot y)} &\leq 2k\cdot 2^{-Q}(\abs{X}\cdot\abs{Y_i})
= 2k\cdot 2^{-Q}\sum_{h=1}^k \left[\abs{x_{r,h}}\abs{y_{i,h}} + \abs{x_{i,h}}\abs{y_{r,h}}\right] \leq 2k\cdot 2^{-Q}(\abs x\cdot \abs y).
\end{align*}
Here we have used Cauchy-Schwarz for the final inequality on both sides. Equation~\eqref{e:jr-dp-bound-cx} immediately follows.
\end{proof}

\section{Implementing 2M ZGEMM}
\label{s:examples}

As concrete examples of Algorithms~\ref{a:2m-ozaki2} and~\ref{a:hybrid-2m-3m}, let us consider complex double-precision GEMM (ZGEMM) with 8-bit integer or floating-point GEMMs underneath. We start by discussing general methodology for choosing moduli.

\subsection{Selecting Modulus Sets}
\label{s:modulus-sets}

Given a minimum accumulator precision $M_{\text{min}}$ dictated by the accuracy requirements and input data, our high-level goal is to find a set of moduli $m_1,\dotsc,m_N$, with product $M\geq M_{\text{min}}$, minimizing runtime. Absolute runtime, of course, is highly sensitive to hardware and implementation details. As a starting point, we may model runtime (for fixed inputs $\mathbf A$, $\mathbf B$, $\sigma$, $\tau$) as some increasing function of $N_r$, the number of real low-precision matrix multiplications%
\footnote{
	For hybrid 2M/3M, this model neglects the differences in pre- and post-processing operation counts between 2M and 3M. In practice, pre- and post-processing are typically implemented in separate memory-bound kernels, and the relatively small operation counts of both 2M and 3M are hidden behind much larger memory or cache latencies. Note that the total memory traffic is an increasing function of $N_r$.
}.%

Under this model, our original goal reduces to maximizing $M$ for a given $N_r$. We assume that moduli will be taken from a predetermined set $\mathcal M$ of candidate moduli, based on the low-precision matrix multiplications used (e.g., $\mathcal M=\{2,\dotsc,256\}$ for 8-bit integers). For 2M Ozaki-II, $\mathcal M$ is further restricted to 2M moduli.

\begin{figure}
\centering
\begin{subfigure}[t]{.49\linewidth}
\includegraphics[width=\linewidth, trim=1mm 0mm 3mm 0mm, clip]{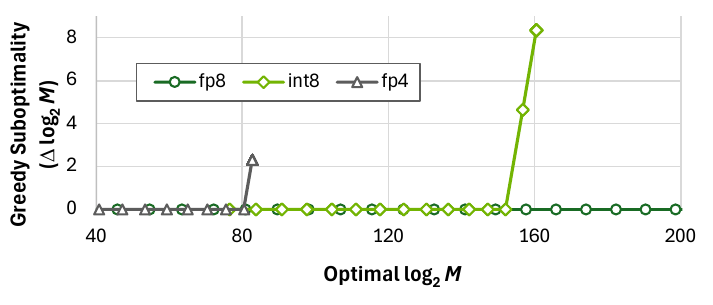}
\caption{Pure 2M (Algorithm~\ref{a:2m-ozaki2})}
\label{f:greedy-subopt-2m}
\end{subfigure}
\hfill
\begin{subfigure}[t]{.49\linewidth}
\includegraphics[width=\linewidth, trim=1mm 0mm 3mm 0mm, clip]{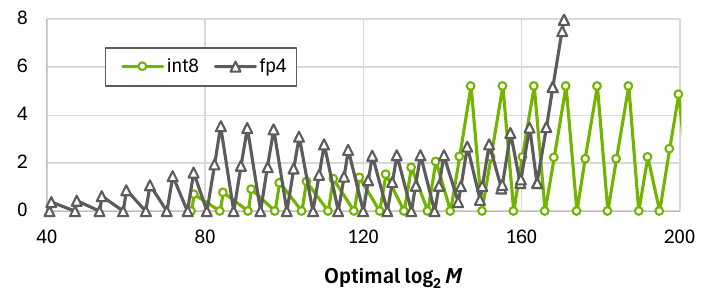}
\caption{Hybrid 2M/3M (Algorithm~\ref{a:hybrid-2m-3m})}
\label{f:greedy-subopt-hybrid}
\end{subfigure}
\caption{Suboptimality of greedy modulus selection. For each optimal set of moduli, a greedy set of moduli is computed with the same maximum cost $N_r$. The gap $\Delta\log_2 M$ between optimal and greedy $\log_2M$ values is displayed.}
\label{f:greedy-subopt}
\end{figure}

For real Ozaki-II, where $N_r = N$, existing implementations generally select moduli via a greedy algorithm~\cite{Ozaki2_2,UchinoFP8}, choosing each $m_\ell$ to be the largest element of $\mathcal M$ coprime to $m_1,\dotsc,m_{\ell-1}$. The same greedy approach extends immediately to 2M Ozaki-II. Alternatively, one may exhaustively search for the optimal modulus set with a branch-and-bound method. Figure~\ref{f:greedy-subopt-2m} compares the $\log_2 M$ values for greedy and exhaustive searches for several data types of interest. In each case, greedy selection is in fact optimal until moduli become scarce.

To extend the greedy algorithm to hybrid 2M/3M Ozaki-II (Algorithm~\ref{a:hybrid-2m-3m}), we take into account the relative efficiency of 2M vs.\ 3M by ranking moduli by \emph{bits per real multiplication}, $r(m) = (\log_2 m) / p(m)$, where $p(m) = 2$ if $m$ is a 2M modulus, and 3 otherwise. Then $m_\ell$ is chosen to be the maximize $r(m)$ among moduli coprime to $m_1,\dotsc,m_{\ell-1}$. As Figure~\ref{f:greedy-subopt-hybrid} shows, greedy selection is optimal for some values of $N_r$ but shows quantization effects that reduce its efficacy at others. A simple example illustrates the issue: consider 8-bit integer moduli with a budget of $N_r=5$. The greedy algorithm selects two 2M moduli $\{241,\, 233\}$, wasting one of the available real multiplications, while the optimal set $\{241,\, 256\}$ uses the additional real multiplication to trade a 2M modulus for a slightly larger non-2M modulus. 

\subsection{ZGEMM with 8-bit Integer GEMMs}
\label{s:zgemm-int8}

\begin{table}[t]
\small
\begin{tabular}{r|ccccc ccccc ccccc}
$\ell$ & 1&2&3&4&5&6&7&8&9&10&11&12&13&14&15\\
\hline
\textbf{2M}&\itshape 241&\itshape 233&\itshape 229&\itshape 221&\itshape 205&\itshape 197&\itshape 193&\itshape 181&\itshape 173&\itshape 157&\itshape 149&\itshape 137&\itshape 113&\itshape 109&\itshape 101\\ 
\textbf{Hybrid}&\multicolumn{15}{c}{\bfseries (same as 2M)}\\
\textbf{3M}&256&255&253&251&247&241&239&233&229&227&223&217&211&199&197
\end{tabular}
\\\bigskip
\begin{tabular}{r|ccccc ccccc ccccc}
$\ell$ &16&17&18&19&20&21&22&23&24&25&26&27&28&29&30\\
\hline
\textbf{2M}&\itshape 97&\itshape 89&\itshape 73&\itshape 61&\itshape 53&\itshape 37&\itshape 29&-&-&-&-&-&-&-&-\\
\textbf{Hybrid}&\multicolumn{5}{c}{\bfseries (same as 2M)}&256&253&251&249&239&227&\itshape 37&223&217&211\\
\textbf{3M}&193&191&181&179&173&167&163&157&151&149&139&137&131&127&113
\end{tabular}
\bigskip
\caption{Greedy int8 moduli sets for 2M, 3M, and hybrid 2M/3M GEMM, with 2M moduli italicized.}
\label{t:int8-greedy}
\end{table}

\begin{figure}
\centering
\begin{subfigure}[t]{.49\linewidth}
\includegraphics[width=\linewidth, trim=0mm 10mm 0mm 2mm, clip]{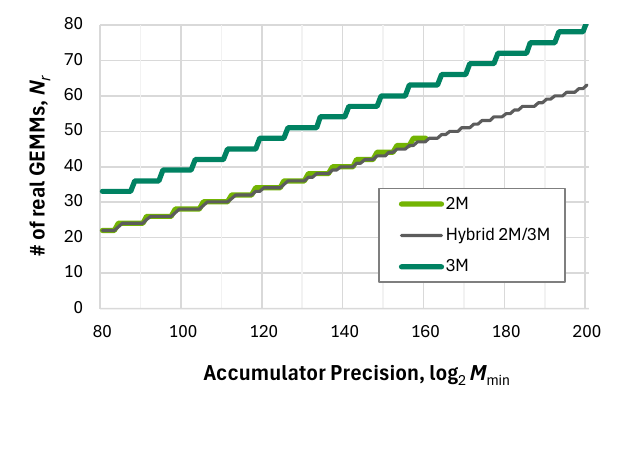}
\caption{Cost in real matrix multiplications, $N_r$.}
\label{f:int8-M-sweep-Nr}
\end{subfigure}
\hfill
\begin{subfigure}[t]{.49\linewidth}
\includegraphics[width=\linewidth, trim=0mm 10mm 0mm 2mm, clip]{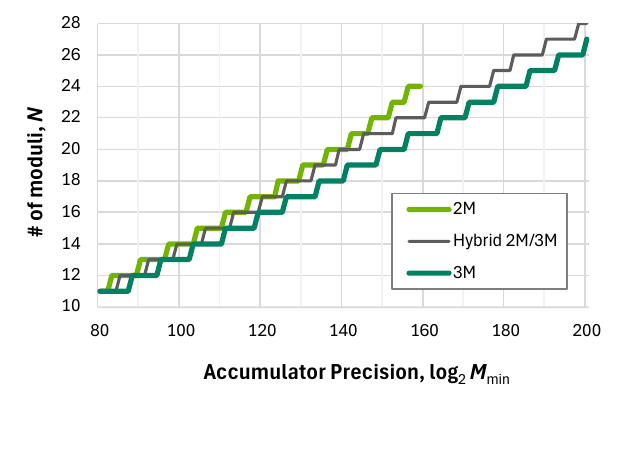}
\caption{Modulus count, $N$.}
\label{f:int8-M-sweep-moduli}
\end{subfigure}
\caption{Modulus count $N$ and real matrix multiplication count $N_r$, as functions of required accumulator precision, $\log_2(M_\text{min})$, for optimal sets of int8 moduli.}
\label{f:int8-M-sweep}
\end{figure}

The original Ozaki-II scheme targets 8-bit integer (int8) matrix multiplication hardware, which accommodates moduli up to 256 without recourse to multiword arithmetic. Table~\ref{t:int8-greedy} lists the modulus sequences generated the greedy method of section~\ref{s:modulus-sets} for the 2M, 3M, and hybrid 2M/3M algorithms.

We notice immediately that the 2M moduli diminish more rapidly than 3M moduli, as expected. The relative sparsity of $\mathcal M_{2M}$ w.r.t. $\mathcal M_{\text{real}}$ reflects the fact that asymptotically only half of the prime numbers are valid 2M moduli, as observed in Section~\ref{s:2m-ozaki2-algebra}. Moreover, after $N=22$ moduli ($\log_2 M\approx 152.1$) the greedy method has exhausted the available 2M moduli, though by exhaustive search it is possible to extend to $N=24$ coprime moduli ($\log_2 M\approx 160.4$). If higher precision is needed (e.g., due to large ESC) we must use hybrid 2M/3M, Algorithm~\ref{a:hybrid-2m-3m}.

Figure~\ref{f:int8-M-sweep} compares the cost of the different algorithms as we vary the required precision $M_{\text{min}}$. Because of the sparsity of 2M moduli, the new algorithms may require an additional 1--2 moduli for a given precision compared to 3M, but nevertheless maintain a significant reduction in overall cost. Note that Figure~\ref{f:int8-M-sweep} is computed using \emph{optimal} modulus sets rather than the greedy choices from Table~\ref{t:int8-greedy}; however, in all cases greedy selection requires at most one additional modulus compared to the optimum.

As mentioned in Section~\ref{s:2m-ozaki2-algebra}, the set of 2M moduli may be varied by discretizing to different imaginary quadratic number rings. Indeed, for 8-bit integer moduli, the 2M moduli for the Eisenstein ring $\ZZ[e^{2\pi i/3}]$ are marginally denser than 2M moduli in $\ZZ[i]$, achieving $\log_2M\approx 167.4$ before exhausting 2M moduli (versus~160.4 for $\ZZ[i]$). Similarly, for hybrid 2M/3M, we may obtain up to approximately $3.7$ additional bits of accumulator precision for the same number of real GEMMs by using $\ZZ[e^{2\pi i/3}]$ in Algorithm~\ref{a:hybrid-2m-3m}. However, this advantage is offset by slightly more complicated quantization/dequantization steps; in particular, quantization involves rounding each input in the complex plane to the nearest point on a hexagonal lattice.

\subsection{ZGEMM with 4-bit and 8-bit Floating-Point GEMMs}

On some GPUs, such as the NVIDIA B300, low-precision floating-point GEMMs are much faster in hardware than integer GEMMs, leading to the development of fp8~\cite{UchinoFP8} and fp4~\cite{HayashiFP4} variants of Ozaki-II, which use the E4M3 and E2M1 formats~\cite{OCPMX,OFP8}, respectively. Modular integer matrices are embedded in the target floating-point format (possibly rescaled), and the floating-point results converted back to integers, carefully avoiding loss of accuracy due to rounding.

Because of the narrower mantissa fields in fp4 and fp8 formats, moduli are restricted to a much smaller range when using single-word arithmetic--- for E2M1, $m_\ell\in\{2,\dotsc,11,13\}$, while for E4M3, Uchino et al.\ limit%
\footnote{
	In fact, single-word E4M3 multiplication can be extended to certain moduli larger than 33, even though $[-16,16]$ is the largest consecutive integer range representable in E4M3. This is because it suffices to find E4M3 representatives of every residue class modulo $m_\ell$, as Hayashi et al.~\cite{HayashiFP4} already leverage for E2M1. Extending E4M3 to larger moduli has some additional complications that are outside the scope of this paper; for simplicity, we follow Uchino et al.'s modulus restrictions here.
}
$m_\ell\leq 33$. As the available moduli are not enough to obtain even single-precision accuracy, existing work uses multi-word arithmetic (multiple fp4/fp8 matrix multiplications per modulus) to accommodate larger moduli. More specifically, for each modulus $m_\ell$ we choose a base $b_\ell \bmod m_\ell$ and split $A$ and $B$ so that
\begin{align*}
A&\Eq A_\ell^{(1)}b_\ell^{\vphantom{(1)}} + A_\ell^{(0)}\pmod{m_\ell},
&
B&\Eq B_\ell^{(1)}b_\ell^{\vphantom{(1)}}  + B_\ell^{(0)}\pmod{m_\ell},
\end{align*}
where $A_\ell^{(i)}$, $A_\ell^{(i)}$ are representable in fp4/fp8. We can then compute $AB\bmod m_\ell$ with 4 fp4/fp8 multiplies modulo $m_\ell$. For certain choices of $m_\ell$ it is possible to use either Karatsuba multiplication or square-optimized multiplication to compute $AB$ with only 3 multiplies, as detailed in~\cite{UchinoFP8}.


Both 2M and hybrid 2M/3M Ozaki-II carry over to this setting. We first apply 2M (or 3M) to reduce the problem of computing a complex product $AB\bmod m$ to 2 or 3 real products $A_qB_q\bmod m$, then apply any double-word multiplication algorithm (schoolbook, Karatsuba, or square-optimized) to reduce each of these real products to a set of fp4 or fp8 matrix multiplications.

With Uchino et al.'s double-word fp8 algorithm, the maximum modulus $m=1089$ is relatively large, allowing 2M Ozaki-II to attain sufficiently high precision ($\log_2 M\approx 350.2$) that hybrid 2M/3M is practically unnecessary. Hayashi et al.'s double-word fp4 has a much more limited range, with maximum modulus $m=169$. Pure 2M can only achieve $\log_2 M\approx 82.7$, so hybrid 2M/3M is always necessary to obtain double precision accuracy.

\section{2M Rank-$k$ Updates%
\protect\footnote{The author would like to thank Robert Parrish for pointing out this method of reducing SYRK to 2M GEMM.}%
}

The 2M method also yields computationally efficient methods for both real and complex BLAS-3 rank-$k$ update routines (\SYRK, \HERK). We recall these routines' definitions, omitting alpha and beta scaling for simplicity:

\begin{itemize}
	\item \SYRK{}: $C=AA\trp$,\; $A\in\RR^{n\times k}$.
	\item \HERK{}: $C=AA\htrp$,\; $A\in\CC^{n\times k}$.
\end{itemize}

In each case $C$ is (Hermitian) symmetric, so only the lower or upper triangle of $C$ is computed and stored.
In theory, the symmetry of $C$ halves the operation count (e.g., $n^2k+o(n^2k)$ flops for SYRK vs. $2n^2k$ for a same-sized GEMM), but library implementations generally cannot obtain the full $2\times$ speedup. The reason is that SYRK-like routines are typically implemented with a modified GEMM kernel that masks its output to the desired triangle of the output matrix $C$~\cite{GotoBLAS3}. As each kernel operates on a fixed-size tile of $C$, there is wasted computation near the diagonal, where tiles lie partly outside the desired triangle. Shrinking tile size reduces waste, but decreases arithmetic intensity, reducing performance.

With 2M, it is possible to implement symmetric products using full, unmasked GEMMs, while still maintaining the expected 2x speedup. The key idea is that Hermitian symmetry in the problem induces a symmetry relation in the real integer products $C_\pm$, allowing us to compute only half as many integer GEMMs. We outline the methods below, which may be applied in either the integer domain (modulo $m$) or in floating point, in conjunction with Ozaki-II.

\paragraph{\HERK} Writing $A=A_r+iA_i$ and applying 2M, we have
\begin{align*}
	C_+ &= (A_r+sA_i)(A_r-sA_i)\trp \bmod m,\\
	C_- &= (A_r-sA_i)(A_r+sA_i)\trp \bmod m.
\end{align*}
In particular $C_-=C_+\trp$, so we only need to compute a single real multiplication $C_+$ for each $m$, providing the expected 2$\times$ speedup while using only full GEMMs.%
\footnote{
	Dumas et al.~\cite{Dumas} present a separate 2M HERK algorithm that reduces $AA^H$ to one real SYRK and one real GEMM, over an arbitrary commutative ring. Asymptotically, their algorithm requires $\frac32$ as many operations, but allows arbitrary moduli.
}

\paragraph{Real \SYRK} We first reduce an arbitrary (real) $n\times k$ \SYRK{} $AA\trp$ to an $n\times (k/2)$ \HERK{} $ZZ\htrp$. We may assume $k$ is even by zero-padding if necessary. Split $A$ evenly in the $k$ dimension: $A=\big[\,A_1\,\big|\, A_2\,\big]$, and form $Z=A_1+iA_2\in\CC^{n\times k/2}$. Now
\begin{align*}
ZZ\htrp &= (A_1A_1\trp+A_2A_2\trp) + i(A_2A_1\trp-A_1A_2\trp)
	= AA\trp + i(A_2A_1\trp-A_1A_2\trp),
\end{align*}
so that $AA\trp=\RRe(ZZ\htrp)$. We compute $ZZ\htrp$ with the 2M \HERK{} algorithm above, discarding the imaginary part.

For the new 2M-based \SYRK{}/\HERK{} algorithms, we pay $\Theta(nk)$ additional pre-processing operations and $\Theta(n^2)$ post-processing operations, compared to the standard masked-GEMM approach, which itself wastes $\Theta(nk\cdot\max(t_R, t_C))$ operations, where $t_R\times t_C$ is the kernel tile size. Hence the benefits of 2M depend on the problem size, in addition to machine- and implementation-dependent details.
In addition, when using Ozaki-II, the 2M SYRK algorithm
may increase the number of required moduli, particularly at higher accumulator precisions, as shown earlier in Figure~\ref{f:int8-M-sweep-moduli}.

\section{Computational Results}

To demonstrate the efficacy of the 2M method, we present performance and accuracy results for int8-emulated 2M ZGEMM on a single NVIDIA B200 GPU. All results are with $\alpha=1$, $\beta = 0$, and use the Gaussian integer version of 2M.

\paragraph{Accuracy Metric} Accuracy is measured as maximum elementwise relative error, compared against a reference quadruple-precision implementation. Real inputs are taken from Uchino et al.'s data distribution~\cite{Ozaki2_2}: $\textbf a_{ih} = u_{ih}\exp(\phi v_{ih})$, with $u_{ih}$ uniformly drawn from $[-\frac12,\frac12)$ and $v_{ih}\sim \mathcal N(0,1)$, and similarly for $\textbf B$. The parameter $\phi>0$ is a difficulty parameter, with higher $\phi$ values producing wider exponent ranges and larger ESC values. Complex input data follows a similar distribution $\textbf a_{ih} = \frac12 u_{ih}\exp(\phi v_{ih}+ 2\pi iw_{ih})$, where $v_{ih}$ is as before, and $u_{ih}$, $w_{ih}$ are uniformly drawn from $[0,1)$.

\paragraph{Performance Metric} Performance of all implementations is measured in effective teraflops per second (TFLOP/s), calculated using the floating point operation count of the \emph{baseline algorithm}: i.e., $8mnk$ operations for ZGEMM and $2mnk$ operations for DGEMM. Effective TFLOP/s numbers do not reflect actual operation counts for Ozaki-II GEMMs (and do not indicate any accuracy guarantees), but provide a standard metric for comparing implementations.

\paragraph{Algorithms} We benchmark an internal implementation of Algorithm~\ref{a:hybrid-2m-3m} (hybrid 2M/3M) with a greedy modulus set. Note that this is identical to Algorithm~\ref{a:2m-ozaki2} for $N\leq 20$ moduli. For comparison, we also test Uchino et al.'s 3M-based \texttt{GEMMul8} library%
	\footnote{Tested version: \url{https://github.com/RIKEN-RCCS/GEMMul8/commit/3115e709}}
as well as an internal implementation of the 3M Ozaki-II algorithm. \texttt{cublasZgemm} from cuBLAS 13.1, with emulation disabled, is used as the reference 4M floating-point GEMM. Scaling factors are determined using fixed mantissa bit count scaling (equation~\eqref{e:standard-sf}) for accuracy tests, and Uchino et al.'s fast scaling method~\cite{Ozaki2_2} for performance tests.

\subsection{Accuracy}

\begin{figure}
\centering
\begin{subfigure}[t]{.49\linewidth}
\includegraphics[width=\linewidth]{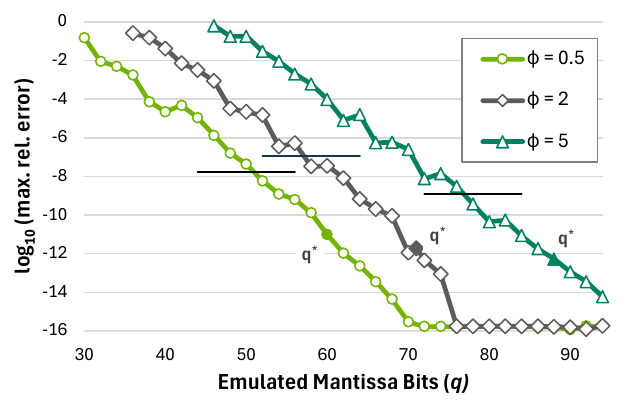}
\caption{DGEMM}
\label{f:accuracy-dgemm}
\end{subfigure}
\hfill
\begin{subfigure}[t]{.46\linewidth}
\includegraphics[width=\linewidth]{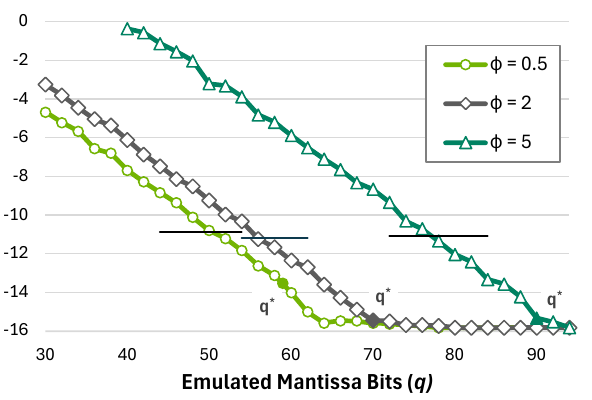}
\caption{Hybrid 2M/3M ZGEMM}
\label{f:accuracy-zgemm}
\end{subfigure}
\caption{Maximum relative error of Ozaki-II DGEMM and hybrid 2M/3M Ozaki-II ZGEMM for input data of varying difficulties $\phi$, as a function of mantissa bits, $q$. Horizontal lines show the accuracy attained by standard floating-point DGEMM/ZGEMM for each choice of $\phi$, while $q^*$ marks the value of $q$ suggested by ESC (Proposition~\ref{p:esc-cx}).
}
\label{f:accuracy}
\end{figure}

We begin by computationally verifying accuracy for 2M and hybrid 2M/3M Ozaki-II ZGEMM. Figure~\ref{f:accuracy-zgemm} plots maximum relative error as a function of mantissa bit count $q$ for several choices of difficulty parameter $\phi\in\{0.5,2,5\}$, for matrices of size $m = n = k = 4096$. The modulus count $N$ is deduced using the worst-case bound $M\geq 2^{2q+2}k$, ranging from $N=10$ at $q=30$ to $N=28$ at $q=94$.

The roughly linear slope in the log-plot reflect the expected $O(2^{-q})$ error decay~(equation~\eqref{e:fxp-fwd-error-cx}), and is qualitatively consistent%
\footnote{
	The large maximum relative errors seen in Figure~\ref{f:accuracy-dgemm} as compared to Figure~\ref{f:accuracy-zgemm} are due to numerical cancellation, which leads to poorly-conditioned dot products. Ill-conditioning is less likely for random complex data since real and imaginary parts must simultaneously exhibit significant cancellation.
}
with Ozaki-II DGEMM (cf.\ Figure~\ref{f:accuracy-dgemm}).
As $\phi$ increases, the ESC increases, as does the number of mantissa bits required to match floating-point GEMM. For each choice of $\phi$, we also compute the ESC-based mantissa bit count $q^*$ from Fasi et al.~\cite{FMM} and Proposition~\ref{p:esc-cx}; across all test cases $q^*$ consistently overpredicts the required number of mantissa bits.

\subsection{Performance}

\begin{figure}
\centering
\begin{subfigure}[t]{.47\linewidth}
\includegraphics[width=\linewidth]{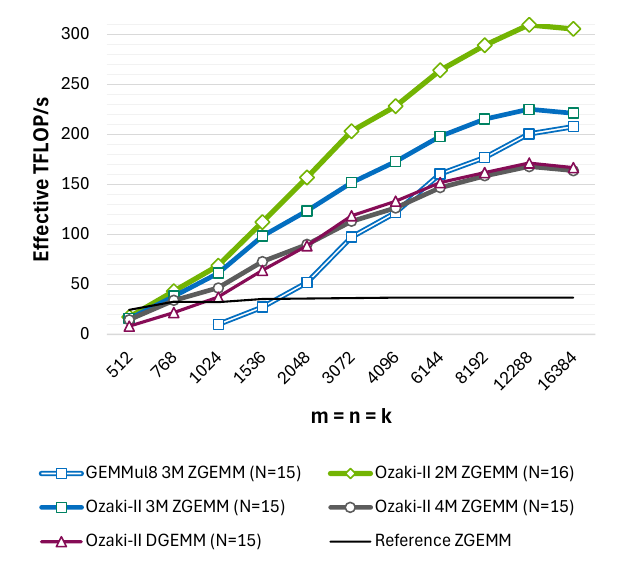}
\caption{Square sizes}
\label{f:perf-sweep-square}
\end{subfigure}
\hfill
\begin{subfigure}[t]{.47\linewidth}
\includegraphics[width=\linewidth]{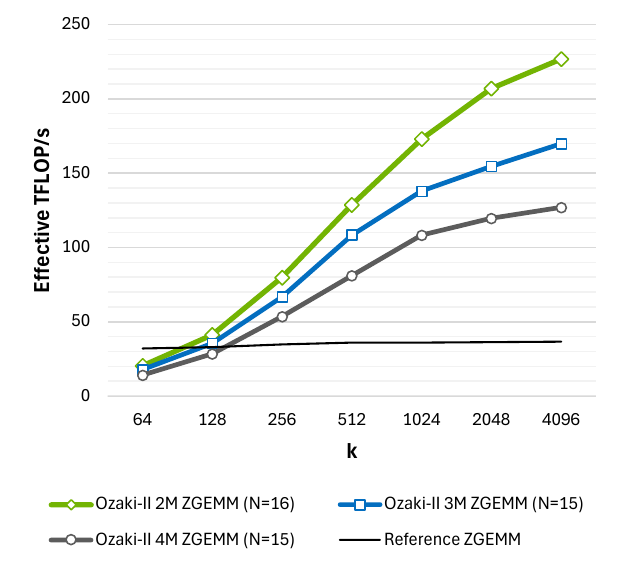}
\caption{Small-$k$ cases, $m = n = 4096$.}
\label{f:perf-sweep-smallk}
\end{subfigure}
\caption{ZGEMM performance benchmarks with fixed accuracy $\log_2 M\approx 117$ and variable sizes, on an NVIDIA B200 card.}
\label{f:perf-sweep}
\end{figure}

Figure~\ref{f:perf-sweep-square} shows 2M, 3M, and 4M Ozaki-II ZGEMM performance across a range of problem sizes, with fixed accumulator precision $\log_2 M\approx 117$. For 2M, $N=16$ moduli are needed to attain this precision; other Ozaki-II algorithms need one fewer modulus, $N=15$. In addition, we compare performance in small-$k$ scenarios (Figure~\ref{f:perf-sweep-smallk}) to make any differences in 2M, 3M, and 4M post-processing cost more prominent. Uchino et al.'s fast scaling algorithm~\cite{Ozaki2_2} was used for all Ozaki-II implementations.

At this precision, 2M ZGEMM is up to $8.4\times$ faster than reference 4M fp64 ZGEMM. It is uniformly faster than 3M and 4M Ozaki-II on all test cases, with speedups up to $1.38\times$ over 3M, and $1.86\times$ over 4M. 2M also achieves an average $1.83\times$ speedup in effective TFLOP/s compared to Ozaki-II DGEMM at the same precision. That is, in absolute time a single ZGEMM costs approximately $4/1.83\approx 2.19$ DGEMMs, approaching the 2-DGEMM theoretical limit. The gap arises primarily because 2M requires an additional modulus to achieve a similar accuracy level (cf.~Figure~\ref{f:int8-M-sweep-moduli}).

\begin{figure}
\centering
\begin{subfigure}[t]{.58\linewidth}
\includegraphics[width=0.98\linewidth, trim=0mm 0mm 0mm 2mm, clip]{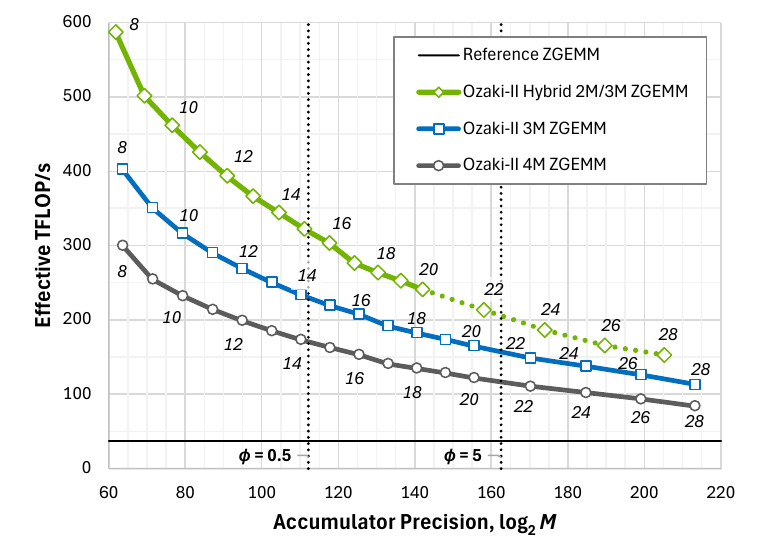}
\caption{Performance vs.\ accumulator precision, $m = n = k = 16384$. Marked data points show modulus count $N$.
Dotted vertical lines mark the precisions required to match reference floating-point GEMM accuracy for $\phi=0.5,\,5$.
} 
\label{f:16k-perf-acc-sweep}
\end{subfigure}
\hfill
\begin{subfigure}[t]{.4\linewidth}
\includegraphics[width=\linewidth, trim=4mm 9mm 9.3mm 6mm, clip]{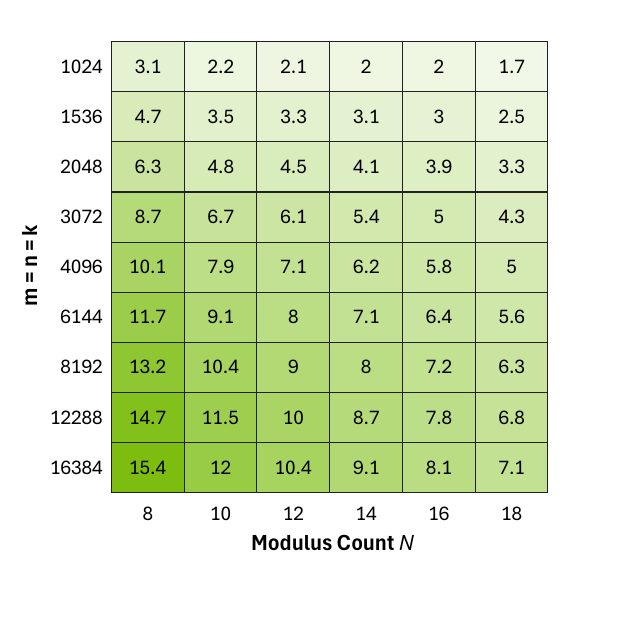}
\caption{Ozaki-II 2M ZGEMM speedups over reference ZGEMM for varying square sizes and modulus counts.}
\label{f:heatmap}
\end{subfigure}
\caption{Additional 2M and hybrid 2M/3M ZGEMM performance data.}
\label{f:results2}
\end{figure}

In Figure~\ref{f:16k-perf-acc-sweep} we fix the problem size to $m=n=k=16384$ and vary $N$ to show the tradeoff between performance and accumulator precision. Hybrid 2M/3M shows similar speedups over 3M ZGEMM and Ozaki-II DGEMM across the range of $N$, with performance up to 586 TFLOP/s (15$\times$ reference ZGEMM) at $N=8$.
Figure~\ref{f:heatmap} provides a more detailed heatmap of speedups over reference ZGEMM across a range of sizes and modulus counts.

\section{Conclusion}

With each successive generation of CPU and GPU hardware, low-precision GEMM throughput continues to outpace that of traditional single and double precision arithmetic. The Ozaki-II scheme and other emulation methods take advantage of this trend, using integer quantization and small integer GEMMs to accelerate floating-point GEMMs.

In this paper, we have shown that this integer-quantized approach opens up new algorithms for complex GEMMs that leverage the algebraic structure of quadratic integer rings. In particular, 2M multiplication reduces a complex GEMM to two real GEMMs of the same size, instead of the three or four real GEMMs required by previous algorithms. For rank-$k$ updates, 2M reduces a complex rank-$k$ update ($C=AA\htrp$) to a real GEMM of the same size, and a real rank-$k$ update ($C=AA\trp$) to a real GEMM of half the size. The new 2M algorithm is easy to implement, and achieves practical speedups that closely match these theoretical predictions.

\subsection*{Acknowledgements}

The author is grateful for helpful reviews from numerous colleagues, which greatly improved this paper.

\bibliographystyle{siam}
\bibliography{2m-gemm}

\end{document}